\documentclass[pdflatex,sn-mathphys-ay]{sn-jnl}

\usepackage{graphicx}%
\usepackage{multirow}%
\usepackage{amsmath,amssymb,amsfonts}%
\usepackage{amsthm}%
\usepackage{mathrsfs}%
\usepackage[title]{appendix}%
\usepackage{xcolor}%
\usepackage{textcomp}%
\usepackage{manyfoot}%
\usepackage{booktabs}%
\usepackage{algorithm}%
\usepackage{algorithmicx}%
\usepackage{algpseudocode}%
\usepackage{listings}%

\theoremstyle{thmstyleone}%
\newtheorem{proposition}{Proposition}%

\theoremstyle{thmstyletwo}%

\theoremstyle{thmstylethree}%

\usepackage{graphicx}
\usepackage{url}
\usepackage{float}
\begin{document}

\title[Executive Judgement in AI-Mediated Decision-Making]{Executive Judgement in AI-Mediated Decision-Making Environments: A Process Theory of Formation, Qualification Attrition and Authorisation}

\author{\fnm{Richard} \sur{Hill}}
\affil{\orgdiv{Department of Computer Science}, \orgname{University of Huddersfield}, \orgaddress{\city{Huddersfield}, \country{UK}}}
\affil{\normalfont\texttt{r.hill@hud.ac.uk}}
\affil{\normalfont ORCID: \href{https://orcid.org/0000-0003-0105-7730}{0000-0003-0105-7730}}

\abstract{Generative artificial intelligence can contribute to the representations, alternatives and evaluations through which executive judgements are formed. Existing research already explains important aspects of hybrid cognition, reliance, managerial agency and accountability. This article develops a narrower process challenge: locally competent human and AI contributions can still culminate in an inadequately warranted organisational commitment when established operational assumptions, uncertainties or dependencies are lost or transformed before authorisation. It proposes qualification attrition as the process through which such epistemically consequential content is weakened across successive transformations, and a formation--authorisation gap as the resulting discrepancy between the grounds required for warranted commitment and those that remain accessible, intelligible and challengeable to the authorising role. Executive judgement governance is defined through four dimensions: boundary setting, interpretive challenge, reliance calibration, and authorisation with answerability. Four propositions connect transformation, qualification attrition and governance arrangements to epistemic risk, accountability risk and judgement quality. The contribution is an integrative process theory of how distributed judgement formation can fail as a composition even where individual contributions, local reliance decisions and formal accountability arrangements appear competent.}


\maketitle

\section{Introduction}

Consider an organisational decision in which each local contribution appears competent. An analyst produces a careful assessment; an AI system summarises it accurately; a manager converts the analysis into an executive briefing; and an authorised executive approves the resulting recommendation. The final commitment can nevertheless be inadequately warranted if concrete assumptions, uncertainty, dependencies or dissent are weakened or lost as the analysis passes between these stages. No single contributor need have made an obvious error. The problem can arise from how otherwise reasonable contributions transform into an authorised commitment.

An executive can authorise an organisational commitment without having produced, or adequately examined, the representations on which it rests. Analysts, professional advisers and information systems have long contributed to this separation. Generative artificial intelligence extends the possible contributors to problem formulation, option generation and evaluation, while increasing the speed and volume with which intermediate representations can be produced and recombined. The central problem is therefore not simply whether humans and AI can jointly reason well. It is whether the consequential grounds of that reasoning survive the transformations through which a distributed process becomes an authorised organisational commitment.

This question does not require a claim that management scholarship has ignored human--AI cognition. Research already theorises conjoined agency, the interdependence of automation and augmentation, and hybrid problem-solving processes (Murray et al., 2021; Raisch and Krakowski, 2021; Raisch and Fomina, 2025). A recent review maps generative AI contributions across multiple components of organisational decision-making (Schulte and Kanbach, 2026). Research on managerial phronesis directly examines how generative AI may affect context-sensitive moral judgement (Lindebaum et al., 2026), while recent organisation theory treats LLMs as reconfiguring organisational knowing, validation and epistemic agency (Faraj et al., 2026). Taken together, these contributions make it difficult to sustain any broad claim that AI participation in judgement remains theoretically unrecognised.

This article develops a narrower problem: the relationship between distributed judgement formation and consequential authorisation. Analysis, interpretation and framing may involve many contributors, while permission to authorise a commitment remains attached to a smaller set of organisational roles or to previously authorised policies. The theoretical issue is not distribution itself. A highly distributed process can preserve excellent epistemic access, while a simple process can conceal a decisive assumption. The issue is whether assumptions, uncertainties and dependencies remain accessible, intelligible and challengeable when provisional analysis is converted into commitment.

This article proposes executive judgement governance as the capability to configure, challenge, calibrate and render answerable the processes through which distributed contributions become organisational commitments. Its central mechanism is the preservation of epistemically consequential qualifications across transformations in the decision process. The article therefore introduces two linked constructs: qualification attrition, describing the loss or weakening of consequential qualifications as representations are transformed, and the formation--authorisation gap, describing the discrepancy this can create between the grounds required for warranted commitment and those available to the authorising role. The focus is the governance process rather than a new executive personality type.

The central argument is that executive value in these settings lies primarily in making and governing warranted commitments, with AI proficiency and analytical performance contributing to that work. This is a normative and theoretical position about the executive role, not an empirically established ranking of the economic value of different skills. It does not imply that optimisation is intrinsically undesirable or that humans necessarily outperform AI. Indeed, the conditional performance of human--AI combinations makes it necessary to investigate when executive intervention adds value and when it obstructs effective delegation (Vaccaro et al., 2024).

\section{Conceptual approach and scope}

This article uses an integrative conceptual synthesis rather than a systematic review or an empirical study. Foundational theories were selected for their relevance to bounded rationality, distributed cognition, interpretation, expertise and responsibility. A targeted review searched scholarly publications and publisher and institutional repositories for human--AI collaboration, generative AI in strategic decisions, reliance and cognitive offloading, organisational learning, organisational knowing, epistemic governance, accountability and agent oversight. This search identified relevant literature available prior to 29th September 2026, with particular attention to publications from 2022 onwards. It prioritised peer-reviewed research and distinguished journal articles, reviews, experiments, conceptual arguments, workshop evidence and relevant working papers.

Selection was directed by the theoretical problem and included sources that challenge, rather than merely support, the proposed contribution. The search was not an exhaustive database census, and no claim is made for comprehensive coverage, saturation or a reproducible systematic-review sample. This reflects the pragmatic demands of research in a rapidly developing landscape. Findings from laboratory tasks and particular occupations are used within their observed scope. Their implications for executive work are identified as theoretical extensions.

The focal unit of analysis is a consequential decision episode embedded in an organisational process: the period in which a problem is framed, possible responses are evaluated, an authorised commitment is made and its grounds can subsequently be reconsidered. The executive role is a position with delegated authority to commit organisational resources or approve policies that do so. A board or executive team may enact that role collectively; an individual can perform some of the relevant cognitive work without possessing executive authority.

Judgement formation means developing an appraisal of what is happening, what matters and what action is warranted. A decision is the selection or authorisation that commits action; it need not await complete agreement among contributors. Cognition refers here to the production, transformation and use of representations. Authority concerns permission to commit; accountability concerns an obligation to provide an account to a forum able to question and evaluate conduct. These distinctions prevent a distributed production process from being mistaken for either a collective consciousness or a complete allocation of responsibility.

\section{Foundations and the contemporary theoretical problem}

\subsection{Judgement under constraint}

Simon's (1955) account of bounded rationality explains why choice is conditioned by limits on knowledge and computation. It does not establish that optimisation is always impossible or inappropriate. AI may relax particular constraints while leaving others, including the specification of objectives and the evaluation of evidence, unresolved. The inference developed here is that cheaper production of analysis can increase the relative importance of deciding what deserves attention and checking whether the analysis addresses the relevant problem. This extends bounded rationality to a particular organisational configuration; it is not a finding reported by Simon.

Executive work also entails translating appraisal into commitments for which someone must answer. Drucker (1967) offers a classic practice-oriented account of executive effectiveness, while Eisenhardt and Zbaracki (1992) show why strategic decision-making cannot be understood through a rational-choice account alone: political and organisational processes also matter. These foundations support attention to whose objectives define an AI-assisted evaluation and who can challenge its premises. Neither establishes that executive judgement is wholly non-computational or that every relevant task must remain with a human.

Naturalistic decision-making adds an account of how experienced practitioners recognise situations and mentally simulate feasible action (Klein, 1998, 2015). Expertise is not a general immunity to error. Tversky and Kahneman (1974, 1981) establish mechanisms through which human estimates and choices can be affected by heuristics and framing. These traditions place complementary constraints on the proposed model. Contextual experience can supply information absent from a model, but an executive's confidence or intuition does not by itself justify rejecting an accurate recommendation. What matters is whether there are grounds for reliance in the particular situation.

Practical wisdom adds a normative dimension to these cognitive accounts. Nonaka and Toyama (2007) already conceptualise strategic management through distributed practical wisdom: judgement about appropriate action need not reside exclusively in a single leader. This foundation supports examining how organisational purposes and particular circumstances enter a commitment. It does not establish that all responsibility is concentrated in executives, or that the normative adequacy of an AI-assisted decision follows from its human approval.

\subsection{Functional participation in cognition}

Distributed cognition already locates cognitive activity across people and artefacts (Hutchins, 1995). Sociomaterial research likewise challenges accounts in which technology is a neutral addition to otherwise independent social activity (Orlikowski, 2007). The present argument builds on these insights rather than treating them as theories that recognised only passive tools. Conjoined agency provides a further account of how human and technological contributions combine in organisational action (Murray et al., 2021).

AI participates functionally in cognition when its outputs affect the representations through which a decision is formed. A system may select evidence, summarise disagreement, propose a causal explanation or generate alternatives. These operations can affect salience and comparison even if the system has no subjective understanding of the situation. Functional participation does not entail consciousness, intentionality, moral agency or an independent entitlement to authorise action. The argument concerns observable effects on a process, not claims about machine experience.

Weick (1995) supplies an account of the social construction of actionable meaning. Contemporary research identifies generative AI roles across attention, analysis and other decision components (Schulte and Kanbach, 2026). Read together, these sources support investigating how generated representations enter organisational sensemaking. They do not show that an LLM engages in sensemaking in every human or phenomenological sense. Nor is all AI use persistent: a one-off drafting request, a recurring advisory workflow and an agent that executes a delegated sequence represent different degrees and forms of participation.

\subsection{Performance and warranted reliance}

Human--AI collaboration does not guarantee complementarity. Vaccaro et al. (2024) synthesise experimental comparisons and find that combinations, on average, do not outperform the better of the human-only and AI-only alternatives, although effects vary across tasks. This finding makes the comparator important: improving on unaided humans is different from improving on the strongest available arrangement. It does not establish how a particular executive team will perform.

Algorithm aversion and appreciation also concern different conditions of advice use. Dietvorst et al. (2015) show avoidance of algorithms after observed error; Logg et al. (2019) show greater weighting of advice attributed to algorithms in their experimental settings. Neither warrants a general claim that executives always defer to AI or reject it. Trust is an attitude, whereas reliance is behaviour; appropriate reliance requires a relationship between that behaviour and the system's suitability for the task (Lee and See, 2004). The familiar distinctions between use, misuse and disuse remain relevant (Parasuraman and Riley, 1997).

Generative AI extends the range of outputs subject to this problem. Noy and Zhang (2023) report benefits in professional writing tasks, and Brynjolfsson et al. (2025) examine productivity in customer support. Such findings establish that useful assistance is possible, but they do not demonstrate improved executive judgement. Conversely, Lee et al. (2025) report associations between confidence in generative AI and self-reported critical thinking among knowledge workers; this survey does not establish causal cognitive decline. The appropriate theoretical question is how organisations preserve an effective capacity to evaluate assistance while obtaining its benefits.

\subsection{From adjacent theories to a narrower contribution}

These developments leave little basis for broad claims of theoretical absence. Distributed human--AI activity is addressed by conjoined agency and hybrid problem-solving; AI contributions to organisational decision processes are explicitly mapped in recent work. Organisational-learning and managerial-phronesis research explains how substitution, time pressure, accountability and reflexivity can alter the development and exercise of judgement. Recent organisation scholarship further argues that LLMs reconfigure data flows, interpretive practices and validation in organisational knowing (Faraj et al., 2026), while Kim et al. (2026) identify epistemic boundedness where decision makers cannot readily verify the veracity and foundations of available information. A recent working paper by McCarthy et al. (2026) likewise theorises how GenAI-generated claims can travel through organisational knowledge infrastructures and accumulate authority before their epistemic status is settled. Accountability scholarship, including moral-crumple-zone analysis and recent agentic-AI work, also shows that formal responsibility can be misaligned with knowledge and effective control (Elish, 2019; Korzyński et al., 2026). The remaining question is narrower: whether warrant-relevant qualifications produced during distributed judgement formation survive the transformations through which an organisation reaches authorised commitment.

A further adjacent explanation is the theory of information compression (TIC). Watson et al. (2022) theorise how judgement costs within judgement networks can compress information, defining compression in terms of reduced variation in information generated for decision-making. Qualification attrition is narrower and differently specified. It concerns selective degradation of epistemically consequential qualifications rather than reduction in informational volume or variation as such. A highly compressed executive summary can preserve the assumptions, uncertainty, provenance, dependencies and dissent necessary for warranted commitment; conversely, a lengthy report can retain substantial information while obscuring the conditionality of its central recommendation. Information compression may therefore occur without qualification attrition, and qualification attrition may occur without substantial compression. The proposed construct adds value only if this distinction produces explanatory or predictive leverage beyond TIC.

Epistemic opacity and epistemic boundedness are also related but distinct. Work on opaque AI asks when reliance on outputs can be epistemically warranted despite limited access to a system's internal reasoning (Gazit, 2026). Kim et al. (2026) define epistemic boundedness as the inability to verify the veracity and foundations of available information. Both concern whether an informational contribution can be warranted. The formation--authorisation gap instead concerns whether the organisational authoriser possesses the consequential grounds required for commitment. An explainable system can participate in a process with a large formation--authorisation gap if qualifications are lost downstream; an opaque or epistemically bounded source can sit within a comparatively small gap when validation evidence, limitations, uncertainty and dependencies remain usable at authorisation. The formation--authorisation gap should therefore not be inferred from model opacity or source-level unverifiability alone.

\begin{table}[htbp]
\caption{Positioning against adjacent explanations}\label{tab:1}
\centering
\small
\begin{tabular}{@{}p{0.20\textwidth}p{0.35\textwidth}p{0.39\textwidth}@{}}
\toprule
Literature & What is explained & Focus of the proposed extension \\
\midrule
Distributed cognition and sociomateriality & Cognitive work spans people and artefacts; technology and practice are intertwined & How consequential grounds and qualifications survive transformation into an authorised commitment \\
\addlinespace[2pt]
Conjoined agency and hybrid problem-solving & Human and technological contributions combine through different processes & How conjoined production processes preserve or lose assumptions, uncertainty, dependencies and dissent before commitment \\
\addlinespace[2pt]
Trust and human--AI complementarity & Reliance and performance depend on task, system and interaction & Why appropriate local reliance can coexist with an inadequately warranted organisational commitment after aggregation \\
\addlinespace[2pt]
Organisational learning and managerial phronesis & AI substitution can alter learning and context-sensitive judgement & How decision-process transformations can weaken warranted commitment independently of the development of the individual judge \\
\addlinespace[2pt]
Organisational knowing and epistemic governance & LLMs can reconfigure data flows, interpretive practices and validation; AI-supported information may be difficult to verify, and claims can travel through knowledge infrastructures and accumulate authority (Faraj et al., 2026; Kim et al., 2026; McCarthy et al., 2026) & Whether selective loss of warrant-relevant qualifications across transformations to authorisation produces a measurable role-relative gap \\
\addlinespace[2pt]
Accountability and moral crumple zones & Answerability is relational; formal accountability can be misaligned with knowledge and effective control (Bovens, 2007; Elish, 2019; Korzyński et al., 2026) & How qualification preservation and role-relative epistemic access at authorisation interact with effective control and answerability \\
\addlinespace[2pt]
Information compression and judgement networks & Judgement costs can compress information and reduce informational variation, with consequences for decision quality & Why selective loss of epistemically consequential qualifications is distinct from compression in volume or variation (Watson et al., 2022) \\
\addlinespace[2pt]
AI opacity and epistemic trust & Reliance can be difficult to warrant when AI reasoning is inaccessible or incomprehensible & Why organisational authorisation can be epistemically under-supported even with explainable AI, or adequately supported despite model opacity, depending on preserved grounds and qualifications (Gazit, 2026) \\
\bottomrule
\end{tabular}
\end{table}

The contribution is an integrative process theory of epistemic transmission from judgement formation to organisational commitment. It does not depend on rediscovering distributed cognition, joint agency, organisational knowing, validation or accountability. McCarthy et al.'s (2026) account of claims travelling through knowledge infrastructures and accumulating authority is a particularly close rival to any broad claim about epistemic transmission. The narrower proposition is that an organisational decision can fail compositionally: competent local contributions, appropriate local reliance and formally allocated accountability can coexist with an inadequately warranted commitment because warrant-relevant qualifications do not survive the transformations between formation and authorisation. The proposed constructs add value only if selective qualification attrition and the resulting role-relative formation--authorisation gap provide explanatory or predictive leverage beyond general circulation, validation failure, epistemic boundedness and accountability misalignment.

\section{From distributed formation to authorised commitment}

\subsection{The conditional asymmetry}

A decision may draw on many contributions without distributing authority in the same way. A model produces a forecast; another system converts it into a narrative; analysts select supporting evidence; an executive committee approves expenditure. The contributors influence different premises, yet the committee may receive only the final synthesis. Increasing the number of contributors need not improve independence: different systems may reuse the same data or inherit the same framing. Conversely, a single well-supported analysis may leave a shorter and more inspectable route to commitment.

The asymmetry is between judgement that may be formed across many people and systems, and authorisation and answerability that remain comparatively concentrated. It is not unique to AI. What AI can change is the volume, speed and recombination of contributions, together with the ease of producing apparently complete explanations. Machine-learning systems already exhibit dependencies, feedback loops and maintenance problems that exceed the trained model considered in isolation (Sculley et al., 2015). Applying that insight to executive decisions suggests investigating the dependency structure of the grounds for action rather than merely counting advisers or models.

The relevant separation is epistemic and organisational, not simply hierarchical. An executive may know a model's identity yet be unable to identify which assumptions changed an option ranking. Alternatively, a distant committee may possess reliable summaries, independent checks and a genuine ability to commission further analysis. We define the formation--authorisation gap as the discrepancy between the consequential grounds, assumptions, uncertainties and dependencies relevant to a warranted organisational commitment and those that remain accessible, intelligible and challengeable to the role authorised to make that commitment. The formation--authorisation gap is proposed here as a relational construct rather than an established measure. It can be small in a highly distributed process and large in a seemingly simple one.

Concentration can occur at different points. For automated operational decisions, authority may have been exercised earlier when an executive approved objectives, thresholds and delegated permissions. Governance then concerns that policy and its exceptions, rather than personal approval of every transaction. The model therefore accommodates bounded automation; it does not assume that meaningful governance requires an executive to inspect each output.

\subsection{Qualification attrition as the linking mechanism}

Distributed judgement formation necessarily transforms information. Evidence may become analysis; analysis may become a model output; model outputs may be synthesised into a recommendation; recommendations may be compressed into an executive paper or encoded in a delegated policy. Compression is not itself a defect. The process becomes epistemically consequential when a transformation weakens or removes a qualification that matters to whether the resulting commitment is warranted.

This process is referred to as qualification attrition: the loss, attenuation or obscuring of consequential assumptions, uncertainty, provenance, dependencies, dissent or conditionality as representations pass through successive stages of a decision process. For example, `Option A performs best if demand remains above X and supplier Y remains available; confidence is moderate' can become `analysis identifies Option A as optimal'. No contributor need be individually incompetent for this transformation to occur. Each local act of summarisation may be reasonable while their composition produces an overqualified organisational conclusion.

This definition distinguishes qualification attrition from generic information loss or compression. Its object is not the amount of information preserved but whether content that conditions the warrant for commitment survives. Compression can be benign when consequential qualifications are retained; qualification attrition can be severe even in information-rich representations when uncertainty or conditionality is selectively weakened. In causal terms, qualification attrition describes the transformation process; the formation--authorisation gap describes the resulting relational epistemic state at commitment; and epistemic risk is a possible consequence, rather than part of either construct's definition.

Qualification attrition provides the proposed mechanism connecting distributed formation to the formation--authorisation gap. Distribution is therefore neither necessary nor sufficient for the gap. What matters is whether epistemically consequential content survives transformation in a form that the authorising role can understand, contest and act upon. AI can intensify this mechanism by increasing the number, speed and apparent completeness of synthetic intermediary representations, but the mechanism predates AI and should be compared empirically with conventional organisational decision processes.

\subsection{Responsibility is not a single residual obligation}

Bovens (2007) treats accountability as a relationship between an actor and a forum, involving explanation and evaluation. This avoids treating accountability as a substance that simply remains with the human after automation. The relevant questions are who owes an account, to whom, for which conduct and with what consequences. Moral responsibility may depend on knowledge, control and the appropriateness of conduct. Managerial accountability concerns organisational expectations attached to a role. Organisational responsibility can attach to collective policies and practices. Professional accountability concerns obligations within a profession. Legal liability and regulatory accountability are institution- and jurisdiction-dependent; this article does not derive their allocation from a general theory of cognition.

Authority to authorise action is analytically separate from these relationships. A person may have authority but limited information, contribute expertise without authority, or be held accountable for a process whose design they cannot alter. Elish's (2019) analysis of moral crumple zones warns against attributing responsibility to the visible human operator where effective control lies elsewhere. An executive-centred model must therefore examine the distribution of control across boards, suppliers, specialist teams and frontline roles, rather than offer a rationale for concentrating blame.

The normative premise adopted here is that assigning a system a cognitive task does not by itself discharge the obligations attached to a human or institutional role. Producing an explanation on request is not equivalent to occupying an accountable organisational position, accepting evaluation or providing remedy. This premise does not settle philosophical debates about possible future artificial moral agents. It specifies the institutional setting examined by the model and leaves open the possibility that responsibility is shared among multiple human and organisational actors.

\subsection{Why the gap matters for judgement quality}

Epistemic risk is the possibility of treating inadequately supported, contextually inappropriate or misleading claims as sufficient grounds for action. Accountability risk is the possibility that an actor cannot provide a defensible account or that responsibility is attributed without an adequate relationship to control. These risks can diverge. An accurate recommendation can emerge from a process whose responsible actors are unclear. A well-documented approval can rest on poor evidence. Neither documentation nor predictive accuracy alone establishes judgement quality.

Judgement quality is defined as the adequacy of an appraisal and commitment relative to the evidence reasonably available, relevant objectives and constraints, consideration of significant alternatives and consequences, and capacity to revise when grounds change. This definition includes epistemic and normative elements; their weighting must be specified for the decision context. It should not be reduced to eventual success, which is affected by luck, nor to procedural compliance, which can coexist with failure. Empirical research should assess outcome quality and justification quality separately before investigating their relationship.

\section{Executive judgement governance}

\subsection{Construct definition and level of analysis}

Executive judgement governance is the capability, enacted through executive roles and organisational routines, to regulate how distributed cognitive contributions are transformed into warranted and answerable commitments. The construct concerns the process connecting cognition to commitment. It is neither a validated category of executive nor a presumption that human intervention improves performance. In AI-mediated settings, executives may enact this capability personally, collectively, or through organisational arrangements that preserve relevant qualifications and intervention rights.

The construct comprises four proposed dimensions: boundary setting, interpretive challenge, reliance calibration and authorisation with answerability. They specify different objects of governance. Boundaries concern admissible tasks and conditions; interpretive challenge concerns problem representations; calibration concerns the evidential grounds for relying on contributions; authorisation and answerability concern commitment, control and explanation. These dimensions may be enacted by different people. The executive capability lies partly in arranging and sustaining their connections, rather than personally performing every check.

The dimensions are proposed as formative components. Together they constitute the capability; they are not interchangeable symptoms of a single underlying personal trait. No claim is made that a reliable psychometric scale or a necessary-and-sufficient taxonomy has been established. The model is weakened if a simpler adjacent construct predicts the relevant outcomes equally well, or if one or more dimensions add no distinguishable mechanism.

\subsection{Boundary setting}

Boundary setting is the ability to specify where AI contributions are admissible, under which assumptions and with which conditions for escalation or withdrawal. Its mechanism is to align delegation with task demands and the evidence available about system performance. The allocation of automation across information acquisition, analysis, selection and action provides an established starting point (Parasuraman et al., 2000). Joint-activity research also emphasises requirements for coordinating and directing automated contributions (Klein et al., 2004).

Failure occurs when permission granted for a bounded use silently becomes authority for a broader one: a summarisation system is treated as an independent verifier, or a historical forecast becomes the basis for a novel market commitment without reconsideration. The expected contribution to judgement quality is reducing unsupported transfer between contexts. Boundary setting is distinct from general risk appetite because it identifies the cognitive function and conditions of use, rather than only the amount of loss an organisation is prepared to tolerate.

Boundaries cannot be specified perfectly in advance. In novel domains, a responsible boundary may permit experimentation while limiting irreversible commitment. Overly restrictive boundaries can exclude useful evidence or preserve poor human practice. The capability therefore interacts with calibration: observed performance should revise the scope of delegation. It also requires an actual route for enforcing a boundary; a policy that cannot interrupt a workflow provides little effective control.

\subsection{Interpretive challenge}

Interpretive challenge is the ability to expose and test the assumptions, exclusions and value choices through which an issue becomes a particular decision problem. Its mechanism is to maintain meaningfully different representations long enough to identify consequential disagreement. Framing research and sensemaking theory explain why representations matter (Tversky and Kahneman, 1981; Weick, 1995). Their application here concerns organisational arrangements for scrutinising AI-mediated representations, not direct evidence from those classical sources about LLMs.

Failure occurs when fluent summaries erase the distinction between evidence and inference, or when several outputs appear to corroborate one another while repeating a shared premise. Doshi and Hauser (2024) show that AI assistance can improve individual creative outputs while reducing collective diversity in a short-story task. This is a bounded illustration of a possible convergence mechanism, not evidence that executive strategy meetings necessarily lose dissent. In the present model, similarity is problematic only when it conceals relevant alternatives or dependencies; agreement may also reflect sound evidence.

Interpretive challenge requires more than inviting a model to argue the opposite case. Generated disagreement can remain dependent on the original model and sources. Relevant challenge may instead require operational evidence, affected stakeholders or an independently developed interpretation. The expected effect is improved problem relevance and detection of omitted consequences. It depends on access to differing knowledge and on organisational permission to contest a preferred narrative. Algorithmic management research makes clear why power and contestation cannot be assumed away (Kellogg et al., 2020).

Interpretive challenge is costly. Endless reframing can prevent timely commitment, especially where delay itself causes harm. Boundary setting should therefore identify when challenge is warranted, and authorisation should determine when unresolved disagreement is acceptable. The construct does not equate maximum dissent with maximum judgement quality.

\subsection{Reliance calibration}

Reliance calibration is the ability to adjust the weight placed on human and AI contributions to evidence of their competence, uncertainty and relevance in the current task. Its mechanism is diagnostic discrimination between contributions that warrant acceptance, checking, modification or rejection. It builds directly on appropriate-reliance research (Lee and See, 2004); its executive extension concerns how such assessments survive aggregation into a recommendation and remain revisable across decision episodes.

Failure includes blanket trust, indiscriminate rejection and checking that confirms only plausibility. A persuasive explanation need not reveal the actual grounds of an output. Sociotechnical analyses also warn that technical evaluations can omit the social context in which a criterion such as fairness has meaning (Mittelstadt et al., 2016; Selbst et al., 2019). Calibration must therefore consider the validity of the evaluation criterion as well as performance against it.

Buçinca et al. (2021) find that cognitive forcing interventions can reduce over-reliance in an experimental decision task, with costs in participants' subjective evaluations. This supports investigating deliberate independent appraisal, but not prescribing friction in every workflow. The expected benefit is selective correction of inappropriate reliance; the costs include time, effort and possible rejection of useful assistance. In executive settings with delayed outcomes, calibration may require near-term checks of assumptions and evidence quality because ultimate strategic success is not yet observable.

Calibration depends on boundary setting to define the relevant task and on interpretive challenge to reveal alternative explanations. It also needs feedback. If decision makers repeatedly accept polished outputs without comparing their grounds with subsequent evidence, the capacity to evaluate them may weaken. Organisational-learning theory identifies why substitution can change how organisations learn, rather than merely how quickly they process a task (Balasubramanian et al., 2022). Longitudinal effects remain a research question for this model.

\subsection{Authorisation and answerability}

Authorisation with answerability is the ability to connect a commitment to identifiable authority, inspectable reasons, effective intervention and a forum entitled to demand an account. Its mechanism is to make the passage from provisional appraisal to organisational commitment explicit. It is more than obtaining a signature: the authorising role must have access to relevant uncertainty and an effective opportunity to alter, limit or refuse the action.

Failure takes several forms. A recommendation can become operational by default; an approver can lack the capacity to question it; or a record can identify who signed without explaining the grounds accepted. Conversely, a richly documented process can obscure where authority actually resides. Its proposed contribution to quality is to connect the epistemic grounds for action with the organisation's capacity to reconsider, correct and explain that action. This dimension differs from broad compliance because it concerns a particular commitment or delegated policy and the reasons sustaining it.

Answerability should be proportionate to influence and control. Responsibility for commissioning a system, maintaining its evidence base, approving its use and acting on its output may belong to different roles. Requiring an executive to reconstruct every model operation would be neither feasible nor necessarily informative. What matters is access to an adequate account of the consequential premises, unresolved limitations and allocation of intervention rights. Where such access cannot be provided, reducing the scope or irreversibility of the commitment may be more defensible than increasing the volume of documentation.

\section{Conceptual model and research propositions}

\subsection{Process relationships}

The model separates five analytically distinct elements: participation and dependency structure, epistemic transformations, qualification attrition, the formation--authorisation gap, and resulting risks and judgement outcomes. Participation varies across cognitive functions, frequency of interaction and discretion to execute actions. These features affect the dependency structure through which representations are produced, but distribution itself is not the proposed failure mechanism. Epistemic transformations concern the selection, summarisation and recombination of representations; qualification attrition concerns whether assumptions, uncertainty, provenance, dependencies, dissent or conditionality are lost, attenuated or obscured through those transformations. The formation--authorisation gap is the resulting role-relative discrepancy in consequential grounds at commitment. Sensemaking concerns how representations become actionable interpretations, while reliance calibration concerns whether confidence supports appropriate reliance.

Executive judgement governance is proposed to influence qualification attrition and, through it, the formation--authorisation gap. Boundary setting limits inappropriate contributions and clarifies conditions of use; interpretive challenge exposes consequential assumptions and preserves alternative representations; calibration tests the evidential weight placed on contributions; authorisation and answerability require the resulting grounds and limitations to remain usable at the point of commitment. Their adequacy affects epistemic and accountability risk, with consequences for judgement quality. These relationships are theoretical propositions rather than estimated effects.

\begin{table}[htbp]
\caption{Relationships in the conceptual model}\label{tab:2}
\centering
\small
\begin{tabular}{@{}p{0.20\textwidth}p{0.35\textwidth}p{0.39\textwidth}@{}}
\toprule
Element & Proposed relationship & Conditions and observable implications \\
\midrule
AI participation and dependency structure & Changes who or what supplies consequential premises and how many transformations occur & Trace actual inputs, transformations and reuse; do not infer independence from contributor count \\
\addlinespace[2pt]
Epistemic transformations & Selection, summarisation and recombination alter the representation of consequential grounds and create opportunities for qualification preservation or loss & Trace successive representations and identify what is retained, changed or omitted at each transformation \\
\addlinespace[2pt]
Qualification attrition and formation--authorisation gap & Selective degradation of consequential qualifications across transformations widens the role-relative gap at authorisation & Code preservation of assumptions, uncertainty, provenance, dependencies, dissent and conditionality; assess whether required grounds are accessible, intelligible and challengeable to the authorising role \\
\addlinespace[2pt]
Challenge and calibration & Preserve independently grounded alternatives and discriminate warranted from unwarranted reliance & Assess whether consequential disagreement survives to authorisation, not reported trust or output count alone \\
\addlinespace[2pt]
Authority and answerability & Connect acceptance of grounds to a role able to act and provide an account & Separate nominal approval from effective control \\
\addlinespace[2pt]
Epistemic and accountability risks & Affect the defensibility and revisability of commitments & Measure justification and outcomes separately; account for stakes and uncertainty \\
\addlinespace[2pt]
Decision feedback & Revises boundaries, assumptions and reliance in later episodes & Delayed or misleading feedback may weaken learning \\
\bottomrule
\end{tabular}
\end{table}

\begin{figure}[htbp]

\centering

\includegraphics[width=0.98\textwidth]{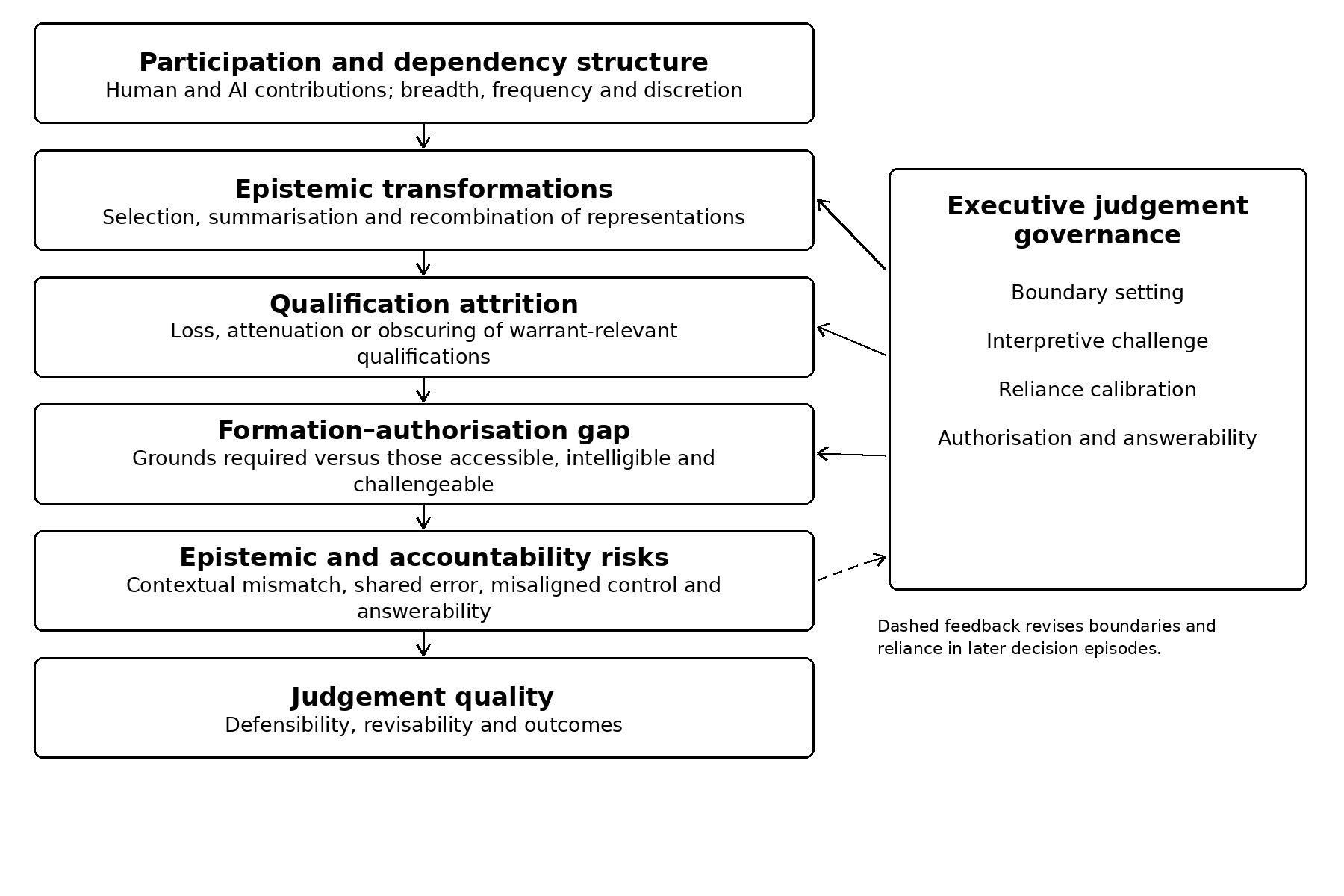}

\caption{Proposed process relationships}\label{fig:process}

\end{figure}

Figure 1 separates epistemic transformations from qualification attrition so that the proposed causal ordering is explicit. Solid arrows show proposed within-episode process relationships. Executive judgement governance acts on transformation conditions, qualification preservation, epistemic access and intervention rights; the dashed feedback relationship revises later decision episodes. The causal relationships shown are proposed rather than empirically estimated in this article.

\subsection{Propositions}

P1 identifies qualification attrition as the central process mechanism. Successive transformations are often necessary, but each creates an opportunity for consequential assumptions, uncertainty, provenance, dependencies or dissent to disappear. Local competence does not prevent this compositional failure: an analyst can summarise reasonably, an AI system can perform accurately and an executive can follow the information presented, while the resulting commitment is nevertheless supported by a thinner set of grounds than the process originally contained.

\begin{proposition}\label{prop:1}
As consequential grounds undergo successive transformations between judgement formation and organisational authorisation, greater qualification attrition increases the formation--authorisation gap, even where individual contributions are locally competent.
\end{proposition}

P2 tests whether the formation--authorisation gap adds explanatory value beyond component performance and local reliance. If the gap merely redescribes inaccurate analysis or inappropriate trust, the proposed theory adds little. The stronger prediction is that epistemic risk can increase because the authorising role lacks consequential qualifications even when upstream analytical performance and local reliance decisions are held constant.

\begin{proposition}\label{prop:2}
Holding task difficulty, analytical performance and the appropriateness of local reliance decisions constant, a larger formation--authorisation gap increases epistemic risk at organisational commitment.
\end{proposition}

P3 concerns preservation through challenge. More outputs do not necessarily preserve qualifications because apparently independent contributions may share data, framing or model dependencies. Independence is not valuable merely as diversity: its proposed mechanism is to expose consequential assumptions, dependencies, uncertainty or dissent that would otherwise be weakened during transformation. Interpretive challenge should therefore reduce qualification attrition only to the extent that such recovered or preserved qualifications remain visible and evaluable at authorisation.

\begin{proposition}\label{prop:3}
Under significant ambiguity, independently grounded interpretive challenge reduces qualification attrition and epistemic risk more than increasing the number of outputs derived from shared premises when it exposes consequential assumptions, dependencies, uncertainty or dissent that would otherwise be lost and preserves them as visible and evaluable at authorisation.
\end{proposition}

P4 connects epistemic access to organisational control. Accountability scholarship already establishes that responsibility can be misaligned with effective control. Recent work on agentic AI goes further: Korzyński et al. (2026) explicitly argue that accountability alignment requires formally accountable decision makers to possess sufficient knowledge, authority, resources and practical control to interrogate, revise or suspend AI-mediated processes. P4 therefore does not claim novelty for the general requirement to align accountability with knowledge and control. Its narrower role is to connect that requirement to the process proposed here: qualification attrition can leave an authorising role unable to exercise otherwise meaningful control intelligently because consequential qualifications no longer survive to authorisation.

P4 therefore connects the manuscript's epistemic process theory with established arguments about accountability and control; it does not propose a new theory of accountability. The manuscript's narrower addition is causal placement: it links the adequacy of epistemic access at commitment to qualification attrition and the formation--authorisation gap, rather than treating knowledge--control alignment as a standalone accountability condition (Bovens, 2007; Elish, 2019; Korzyński et al., 2026).

\begin{proposition}\label{prop:4}
Where qualification attrition has produced a formation--authorisation gap, reducing that gap while preserving effective intervention rights reduces accountability risk more than adding nominal human approval without changing epistemic access or control.
\end{proposition}

The propositions separate process, epistemic and accountability outcomes. P1 concerns the emergence of the formation--authorisation gap through qualification attrition; P2 tests whether that gap predicts epistemic risk beyond component performance and local reliance; P3 specifies a governance mechanism for preserving consequential qualifications; and P4 concerns the alignment of epistemic access, control and answerability. Their integration does not imply that more checking is always beneficial. Preserving every detail would make authorisation impossible; the theoretical problem is preserving consequential qualifications while allowing necessary compression and delegation.

\section{Domain illustrations and evidential limits}

\subsection{Strategic evaluation}

Doshi et al. (2025) examine evaluations of business models in two studies. Individual AI evaluations can be inconsistent or biased, while aggregated evaluations more closely resemble expert rankings. This evidence is directly relevant to strategic evaluation, but agreement with experts is not equivalent to realised strategic success. Nor does it establish that AI reduces dissent in executive meetings or makes an AI-supported commitment more defensible.

The theoretical illustration is a committee receiving a ranking whose apparent stability depends on how models, prompts and roles were aggregated. Boundary setting asks whether the evaluation criterion matches the strategic issue; challenge asks what consequential assumptions were omitted; calibration examines the evidence supporting the ranking; authorisation connects that assessment to the commitment. These implications are an interpretation of the study rather than observed executive behaviours. Shrestha et al. (2019) remains relevant as a conceptual account of decision structures, rather than evidence of premature strategic convergence.

\subsection{Clinical advice and organisational deployment}

Gaube et al. (2021) study practising physicians' responses to chest X-rays with accurate or inaccurate diagnostic advice attributed to an AI system or a human radiologist. The experiment documents susceptibility to incorrect advice across expertise levels and differences in how advice is evaluated. It is not a field demonstration that clinicians abandoned a measured prior intuition, nor does it establish an AI-specific effect in every setting. Its relevance is the possibility that placing a qualified human alongside advice does not ensure that erroneous advice will be rejected.

Zech et al. (2018) separately demonstrate variable generalisation across hospital datasets in pneumonia detection. That study concerns model performance and dataset characteristics, not clinician reliance. Together, these distinct findings motivate examining both contextual validity and the process that authorises use. Kelly et al. (2019) provide a review of the challenges involved in translating AI into clinical impact. The executive-level inference is that organisational deployment decisions require evidence about the system in context and the arrangements for its use; none of these studies validates the proposed executive construct or determines legal liability.

\subsection{Credit decisions}

Fuster et al. (2022) analyse the distributional implications of machine learning in credit markets. Their results do not justify equating better prediction with an unqualified improvement for every borrower group. The illustration concerns the distinction between evaluating a predictor and authorising a policy in which its predictions become grounds for allocating credit. MacKenzie's (2006) historical account of financial models supplies a broader perspective on how models can shape markets; it is not evidence about contemporary generative AI or the later financial crisis.

The present model asks how a financial institution connects predictive evidence, distributional choices and delegated authority. The underlying empirical work does not establish that the four governance dimensions improve credit decisions. It identifies a setting in which predictive and normative criteria can differ, making the proposed distinction between analytical performance and warranted commitment consequential.

\subsection{Agentic workflows}

Agentic configurations extend the problem when a system can take sequences of actions under delegated permissions. Schmitz et al. (2025), in workshop research combining literature review and interviews with civil servants, identify intensified oversight and coordination challenges in public-sector settings. The evidence is emerging and context-specific, rather than a representative test of executive governance across sectors.

For the present model, the implication is that the focal commitment may be the approval of a policy governing a class of actions. Challenge and calibration then operate through evaluation of that policy, monitoring and exceptions. As the time available to intervene decreases, effective governance may require restrictions on permissible action and reversibility established before execution. This is a theoretical implication of the model, not a claim that an executive can continuously supervise every agent action.

\section{Boundary conditions and empirical development}

\subsection{Where the argument applies}

The model is most relevant where AI meaningfully influences consequential premises, objectives or evidence are contested, and authority can be located in identifiable organisational roles. Its marginal value may be smaller for stable, low-stakes tasks with rapid feedback and validated automated procedures. It is not intended to replace technical model assurance or domain expertise. Those capabilities provide essential inputs to the judgements examined here.

The proposed asymmetry is also weaker where authority and knowledge are closely co-located, and different where decisions are genuinely decentralised. Institutional arrangements may assign approval to a board, professional committee or external authority rather than an executive. In such cases the same questions can be asked of the relevant role, but transferring the executive label would obscure the setting. The model assumes that an organisation can exert at least some control over use, even if it cannot inspect or modify a supplier's model. Where that assumption fails, limiting dependence may matter more than refining internal review.

Power is a further boundary condition. Executives may selectively endorse evidence that supports a preferred decision or suppress challenge. Governance routines can legitimise a predetermined commitment rather than improve its grounds. This possibility is consistent with political accounts of strategic decision-making and research on algorithmic control (Eisenhardt and Zbaracki, 1992; Kellogg et al., 2020). The theory therefore assumes neither benevolent executive intentions nor faithful enactment of formal procedures.

\subsection{Operationalisation and research design}

The four dimensions should initially be investigated through observable practices rather than a self-report executive-capability score. Boundary setting can be examined through actual task restrictions, exceptions and revisions. Interpretive challenge can be assessed through whether consequential premises are contested and whether the challenge draws on distinct evidence. Calibration can be evaluated by appropriate acceptance and rejection, including errors introduced by unnecessary overrides. Authorisation and answerability can be studied through the match between formal responsibility, access to grounds and effective intervention.

Qualification attrition and the formation--authorisation gap require independent measurement. Researchers could compare successive representations of the same recommendation and code whether consequential assumptions, uncertainty, provenance, dependencies, dissent and conditionality are preserved, weakened or removed. They could then ask authorising participants to identify the principal grounds and qualifications behind the recommendation and compare those accounts with the documented formation process. The gap should be measured through accessibility, intelligibility and challengeability, not inferred from a bad outcome.

Experiments could hold analytical outputs constant while varying whether consequential qualifications survive into the authorisation representation, as well as provenance, independent challenge and intervention rights. This would separate transformation and governance effects from model capability. Process studies could trace successive drafts of executive recommendations and identify when uncertainty, dissent or dependencies disappear before approval. Longitudinal studies could examine whether feedback changes delegated boundaries and whether independent appraisal is maintained over time. Comparisons should include unaided human processes and conventional organisational summarisation as well as AI-mediated workflows.

Evaluation should include the time and cost of review, the quality of reasons available at commitment, the detection of consequential errors, reversibility, and subsequent outcomes where observable. Normative criteria should be specified before inspecting results and, where relevant, include affected stakeholders. Otherwise researchers may classify decisions as high-quality merely because they conform to the researcher's preferred values or later succeed.

\subsection{Falsification and rival explanations}

The model would be weakened if the proposed dimensions fail to explain variation beyond ordinary decision-process quality, domain expertise, organisational resources and existing AI-governance arrangements. P1 would be challenged if qualification attrition did not predict the formation--authorisation gap once transformation structure and local competence were held constant. P2 would be challenged if the formation--authorisation gap did not increase epistemic risk when task difficulty, analytical performance and the appropriateness of local reliance were controlled. P3 would be challenged if independently grounded interpretive challenge added no preservation or risk-reduction benefit beyond additional outputs derived from shared premises; negative net effects under costly checking remain possible and require prospectively specified boundary conditions. P4 would be weakened if nominal sign-off performed as well as reducing the gap while preserving meaningful intervention rights on independently measured accountability outcomes.

Reverse causality is plausible: difficult or failing decisions may attract more governance, making effective arrangements appear associated with worse outcomes. Well-resourced organisations may both govern better and deploy better models. Strong leaders may improve decisions through mechanisms unrelated to AI. Empirical work must distinguish these explanations. More fundamentally, Faraj et al. (2026), Kim et al. (2026), McCarthy et al. (2026) and Korzyński et al. (2026) already theorise important parts of the surrounding epistemic and governance problem: reconfigured organisational knowing, source-level unverifiability, the circulation and authority of AI-generated claims, and alignment between accountability, knowledge and control. If qualification attrition and the formation--authorisation gap do not explain variation beyond those adjacent mechanisms, the contribution should be treated as a useful synthesis rather than a distinct theoretical advancement.

\section{Conclusion}

The literature establishes that human--AI arrangements can improve some kinds of work, that their benefits depend on task and configuration, and that reliance, learning and responsibility cannot be inferred from output quality alone. It already offers substantial theories of hybrid cognition and managerial judgement, and increasingly theorises organisational knowing, epistemic verification and the alignment of accountability with knowledge and control. The task addressed here is narrower: to specify what happens to warrant-relevant qualifications as these cognitive arrangements are transformed into commitments by authorised organisational roles.

The central difficulty is not distributed cognition itself, but the possibility of compositional epistemic failure between distributed formation and authorised commitment. Qualification attrition can remove consequential assumptions, uncertainty, provenance, dependencies or dissent as representations are transformed. The resulting formation--authorisation gap can leave an authorised role with grounds that appear sufficient while omitting qualifications necessary for warranted commitment. Executive judgement governance addresses this process through boundary setting, interpretive challenge, reliance calibration, and authorisation with answerability.

The argument reframes executive capability around preserving the epistemic adequacy of organisational commitments across transformations in distributed judgement formation. Technical competence matters because it can improve component performance; appropriate reliance matters because it can improve local use of those components. Neither guarantees that consequential qualifications survive aggregation. Equally, preserving a human signature without preserving epistemic access and effective control provides little assurance. The practical implication is to examine the transformation chain between judgement formation and authorisation rather than infer sound governance from adoption, fluency, accuracy or nominal oversight.

Future research should investigate this transformation chain directly, including cases where qualification preservation creates delay, information overload or defensive documentation. The decisive test is whether qualification attrition and the formation--authorisation gap explain variation in epistemic risk beyond analytical performance, appropriate local reliance and ordinary decision-process quality. If they do not, the model should be treated as synthesis rather than distinct theoretical advancement. Until those tests are conducted, the article offers a bounded process theory and research agenda, not a demonstrated solution to executive judgement under AI mediation.

\clearpage
\section*{References}
\begingroup
\setlength{\parindent}{0pt}
\setlength{\parskip}{0.65em}
Balasubramanian, N., Ye, Y. and Xu, M. (2022). Substituting human decision-making with machine learning: Implications for organizational learning. Academy of Management Review, 47(3), 448--465. \url{https://doi.org/10.5465/amr.2019.0470}

Bovens, M. (2007). Analysing and assessing accountability: A conceptual framework. European Law Journal, 13(4), 447--468. \url{https://doi.org/10.1111/j.1468-0386.2007.00378.x}

Brynjolfsson, E., Li, D. and Raymond, L. (2025). Generative AI at work. The Quarterly Journal of Economics, 140(2), 889--942. \url{https://doi.org/10.1093/qje/qjae044}

Buçinca, Z., Malaya, M. B. and Gajos, K. Z. (2021). To trust or to think: Cognitive forcing functions can reduce overreliance on AI in AI-assisted decision-making. Proceedings of the ACM on Human-Computer Interaction, 5(CSCW1), Article 188, 1--21. \url{https://doi.org/10.1145/3449287}

Dietvorst, B. J., Simmons, J. P. and Massey, C. (2015). Algorithm aversion: People erroneously avoid algorithms after seeing them err. Journal of Experimental Psychology: General, 144(1), 114--126. \url{https://doi.org/10.1037/xge0000033}

Doshi, A. R. and Hauser, O. P. (2024). Generative AI enhances individual creativity but reduces the collective diversity of novel content. Science Advances, 10(28), eadn5290. \url{https://doi.org/10.1126/sciadv.adn5290}

Doshi, A. R., Bell, J. J., Mirzayev, E. and Vanneste, B. S. (2025). Generative artificial intelligence and evaluating strategic decisions. Strategic Management Journal, 46(3), 583--610. \url{https://doi.org/10.1002/smj.3677}

Drucker, P. F. (1967). The effective executive. Harper \& Row. \url{https://search.worldcat.org/title/The-effective-executive/oclc/229476}

Eisenhardt, K. M. and Zbaracki, M. J. (1992). Strategic decision making. Strategic Management Journal, 13(S2), 17--37. \url{https://doi.org/10.1002/smj.4250130904}

Elish, M. C. (2019). Moral crumple zones: Cautionary tales in human-robot interaction. Engaging Science, Technology, and Society, 5, 40--60. \url{https://doi.org/10.17351/ests2019.260}

Faraj, S., Perez-Torrents, J., Mantere, S. and Bhardwaj, A. (2026). A time for monsters: Organizational knowing after large language models. Strategic Organization, 24(2), 343--356. \url{https://doi.org/10.1177/14761270251410675}

Fuster, A., Goldsmith-Pinkham, P., Ramadorai, T. and Walther, A. (2022). Predictably unequal? The effects of machine learning on credit markets. The Journal of Finance, 77(1), 5--47. \url{https://doi.org/10.1111/jofi.13090}

Gazit, L. (2026). Constitutive knowledge sources: An institutional approach to epistemic trust in opaque AI systems. AI and Ethics, 6, Article 58. \url{https://doi.org/10.1007/s43681-025-00930-2}

Gaube, S., Suresh, H., Raue, M., Merritt, A., Berkowitz, S. J., Lermer, E., Coughlin, J. F., Guttag, J. V., Colak, E. and Ghassemi, M. (2021). Do as AI say: Susceptibility in deployment of clinical decision-aids. npj Digital Medicine, 4, Article 31. \url{https://doi.org/10.1038/s41746-021-00385-9}

Hutchins, E. (1995). Cognition in the wild. MIT Press. \url{https://mitpress.mit.edu/9780262581462/cognition-in-the-wild/}

Kellogg, K. C., Valentine, M. A. and Christin, A. (2020). Algorithms at work: The new contested terrain of control. Academy of Management Annals, 14(1), 366--410. \url{https://doi.org/10.5465/annals.2018.0174}

Kelly, C. J., Karthikesalingam, A., Suleyman, M., Corrado, G. and King, D. (2019). Key challenges for delivering clinical impact with artificial intelligence. BMC Medicine, 17, Article 195. \url{https://doi.org/10.1186/s12916-019-1426-2}

Kim, Y., Kim, J., Kim, T. and Cho, H.-C. (2026). Administrative decision-making with generative AI: The challenge of epistemic boundedness. Administration \& Society, 58(2), 284--304. \url{https://doi.org/10.1177/00953997251409156}

Klein, G. (1998). Sources of power: How people make decisions. MIT Press. \url{https://mitpress.mit.edu/9780262112277/sources-of-power/}

Klein, G. (2015). A naturalistic decision making perspective on studying intuitive decision making. Journal of Applied Research in Memory and Cognition, 4(3), 164--168. \url{https://doi.org/10.1016/j.jarmac.2015.07.001}

Klein, G., Woods, D. D., Bradshaw, J. M., Hoffman, R. R. and Feltovich, P. J. (2004). Ten challenges for making automation a ``team player'' in joint human-agent activity. IEEE Intelligent Systems, 19(6), 91--95. \url{https://doi.org/10.1109/MIS.2004.74}

Korzyński, P., Wojtczuk-Turek, A., Turek, D., Meglio, O. and Żabicka-Włodarczyk, M. (2026). When agentic AI participates in organizational change: Rethinking agency, leadership, and accountability. Journal of Change Management, 26(3), 203--221. \url{https://doi.org/10.1080/14697017.2026.2719997}

Lee, H.-P., Sarkar, A., Tankelevitch, L., Drosos, I., Rintel, S., Banks, R. and Wilson, N. (2025). The impact of generative AI on critical thinking: Self-reported reductions in cognitive effort and confidence effects from a survey of knowledge workers. Proceedings of the 2025 CHI Conference on Human Factors in Computing Systems, Article 1121, 1--22. ACM. \url{https://doi.org/10.1145/3706598.3713778}

Lee, J. D. and See, K. A. (2004). Trust in automation: Designing for appropriate reliance. Human Factors, 46(1), 50--80. \url{https://doi.org/10.1518/hfes.46.1.50_30392}

Lindebaum, D., Balasubramanian, N., Ashraf, M. and Haack, P. (2026). A process model of managerial phronesis in the age of generative AI. Academy of Management Review. Advance online publication. \url{https://doi.org/10.5465/amr.2024.0582}

Logg, J. M., Minson, J. A. and Moore, D. A. (2019). Algorithm appreciation: People prefer algorithmic to human judgment. Organizational Behavior and Human Decision Processes, 151, 90--103. \url{https://doi.org/10.1016/j.obhdp.2018.12.005}

MacKenzie, D. (2006). An engine, not a camera: How financial models shape markets. MIT Press. \url{https://doi.org/10.7551/mitpress/9780262134606.001.0001}

McCarthy, I. P., Hannigan, T. and Baum, J. A. C. (2026). Generative AI as an epistemic technology: Rethinking governance of organizational knowledge infrastructures. SSRN working paper. \url{https://doi.org/10.2139/ssrn.7390058}

Mittelstadt, B. D., Allo, P., Taddeo, M., Wachter, S. and Floridi, L. (2016). The ethics of algorithms: Mapping the debate. Big Data \& Society, 3(2), 1--21. \url{https://doi.org/10.1177/2053951716679679}

Murray, A., Rhymer, J. and Sirmon, D. G. (2021). Humans and technology: Forms of conjoined agency in organizations. Academy of Management Review, 46(3), 552--571. \url{https://doi.org/10.5465/amr.2019.0186}

Nonaka, I. and Toyama, R. (2007). Strategic management as distributed practical wisdom (phronesis). Industrial and Corporate Change, 16(3), 371--394. \url{https://doi.org/10.1093/icc/dtm014}

Noy, S. and Zhang, W. (2023). Experimental evidence on the productivity effects of generative artificial intelligence. Science, 381(6654), 187--192. \url{https://doi.org/10.1126/science.adh2586}

Orlikowski, W. J. (2007). Sociomaterial practices: Exploring technology at work. Organization Studies, 28(9), 1435--1448. \url{https://doi.org/10.1177/0170840607081138}

Parasuraman, R. and Riley, V. (1997). Humans and automation: Use, misuse, disuse, abuse. Human Factors, 39(2), 230--253. \url{https://doi.org/10.1518/001872097778543886}

Parasuraman, R., Sheridan, T. B. and Wickens, C. D. (2000). A model for types and levels of human interaction with automation. IEEE Transactions on Systems, Man, and Cybernetics, Part A: Systems and Humans, 30(3), 286--297. \url{https://doi.org/10.1109/3468.844354}

Raisch, S. and Fomina, K. (2025). Combining human and artificial intelligence: Hybrid problem-solving in organizations. Academy of Management Review, 50(2), 441--464. \url{https://doi.org/10.5465/amr.2021.0421}

Raisch, S. and Krakowski, S. (2021). Artificial intelligence and management: The automation--augmentation paradox. Academy of Management Review, 46(1), 192--210. \url{https://doi.org/10.5465/amr.2018.0072}

Schmitz, C., Rystrøm, J. and Batzner, J. (2025). Oversight structures for agentic AI in public-sector organizations. Proceedings of the 1st Workshop for Research on Agent Language Models (REALM 2025), 298--308. Association for Computational Linguistics. \url{https://doi.org/10.18653/v1/2025.realm-1.21}

Schulte, N. and Kanbach, D. K. (2026). Rethinking organizational decision-making: The emerging roles and tasks of generative artificial intelligence. Management Review Quarterly. Advance online publication. \url{https://doi.org/10.1007/s11301-026-00611-2}

Sculley, D., Holt, G., Golovin, D., Davydov, E., Phillips, T., Ebner, D., Chaudhary, V., Young, M., Crespo, J.-F. and Dennison, D. (2015). Hidden technical debt in machine learning systems. Advances in Neural Information Processing Systems, 28, 2503--2511. \url{https://research.google/pubs/hidden-technical-debt-in-machine-learning-systems/}

Selbst, A. D., boyd, d., Friedler, S. A., Venkatasubramanian, S. and Vertesi, J. (2019). Fairness and abstraction in sociotechnical systems. Proceedings of the Conference on Fairness, Accountability, and Transparency, 59--68. ACM. \url{https://doi.org/10.1145/3287560.3287598}

Shrestha, Y. R., Ben-Menahem, S. M. and von Krogh, G. (2019). Organizational decision-making structures in the age of artificial intelligence. California Management Review, 61(4), 66--83. \url{https://doi.org/10.1177/0008125619862257}

Simon, H. A. (1955). A behavioral model of rational choice. The Quarterly Journal of Economics, 69(1), 99--118. \url{https://doi.org/10.2307/1884852}

Tversky, A. and Kahneman, D. (1974). Judgment under uncertainty: Heuristics and biases. Science, 185(4157), 1124--1131. \url{https://doi.org/10.1126/science.185.4157.1124}

Tversky, A. and Kahneman, D. (1981). The framing of decisions and the psychology of choice. Science, 211(4481), 453--458. \url{https://doi.org/10.1126/science.7455683}

Vaccaro, M., Almaatouq, A. and Malone, T. (2024). When combinations of humans and AI are useful: A systematic review and meta-analysis. Nature Human Behaviour, 8(12), 2293--2303. \url{https://doi.org/10.1038/s41562-024-02024-1}

Watson, R. T., Plangger, K., Pitt, L. and Tiwana, A. (2022). A theory of information compression: When judgments are costly. Information Systems Research, 34(3), 1089--1108. \url{https://doi.org/10.1287/isre.2022.1163}

Weick, K. E. (1995). Sensemaking in organizations. Sage. \url{https://us2.sagepub.com/en-us/nam/book/sensemaking-organizations}

Zech, J. R., Badgeley, M. A., Liu, M., Costa, A. B., Titano, J. J. and Oermann, E. K. (2018). Variable generalization performance of a deep learning model to detect pneumonia in chest radiographs: A cross-sectional study. PLOS Medicine, 15(11), e1002683. \url{https://doi.org/10.1371/journal.pmed.1002683}
\endgroup

\end{document}